# Symmetric Complementary Logic Inverter Using Integrated Black Phosphorus and MoS$_2$ Transistors

Yang Su, Chaitanya U Kshirsagar, Matthew C Robbins, Nazila Haratipour, and Steven J Koester
*Department of Electrical and Computer Engineering, University of Minnesota, Minneapolis, Minnesota, United States*

**Abstract**

The operation of an integrated two-dimensional complementary metal-oxide-semiconductor inverter with well-matched input/output voltages is reported. The circuit combines a few-layer MoS$_2$ n-MOSFET and a black phosphorus (BP) p-MOSFET fabricated using a common local backgate electrode with thin (20 nm) HfO$_2$ gate dielectric. The constituent devices have linear threshold voltages of −0.8 V and +0.8 V and produce peak transconductances of 16 µS/µm and 41 µS/µm for the MoS$_2$ n-MOSFET and BP p-MOSFET, respectively. The inverter shows a voltage gain of 3.5 at a supply voltage, $V_{DD}$ = 2.5 V, and has peak switching current of 108 µA and off-state current of 8.4 µA (2.4 µA) at $V_{IN}$ = 0 ($V_{IN}$ = 2.5 V). In addition, the inverter has voltage gain greater than unity for $V_{DD} \geq 0.5$ V, has open butterfly curves for $V_{DD} \geq 1$ V, and achieves static noise margin over 500 mV at $V_{DD}$ = 2.5 V. The voltage gain was found to be insensitive to temperature between 270 K and 340 K, and AC large and small-signal operation was demonstrated at frequencies up to 100 kHz. The demonstration of a complementary 2D inverter which operates in a symmetric voltage window suitable for driving a subsequent logic stage is a significant step forward in developing practical applications for devices based upon 2D materials.

Symmetric Complementary Logic Inverter Using Integrated Black Phosphorus and $MoS_2$ Transistors

**1. Introduction**

    Two-dimensional (2D) materials are of interest for numerous device applications because of their layered crystal structure which provides excellent thickness scalability down to the sub-nanometer level, and enables materials integration onto arbitrary substrates.[1-3] The most commonly studied 2D material, graphene, is attractive due to its high carrier mobility, but the absence of an energy gap limits its usefulness for use in logic transistors.[4,5] This shortcoming has led to the exploration of other 2D materials, including transition metal dichalchogenides (TMDs) such as $MoS_2$,[6-8] $MoSe_2$,[9] $MoTe_2$,[10,11] $WSe_2$,[12] $WS_2$,[13] and more recently, black phosphorus (BP).[14-16] These materials are advantageous since they have a layered crystal structure and can be realized in monolayer form, but also have a finite band gap and can thus create high-performance transistors with high on-to-off current ratio. Therefore, TMDs and BP are promising for future scaled CMOS circuits, as well as thin-film and flexible electronics applications. Recent reports have shown excellent performance for $MoS_2$ n-MOSFETs including high drive current and low contact resistance[17,18] and nearly-ideal subthreshold slope.[19] More recently, BP has been shown to be promising for use in high-performance MOSFETs.[20] Despite the promise of 2D semiconductors for use in discrete transistors, building complementary circuits has proven difficult. This is due to the difficulty of fabricating high-performance n- and p-MOSFETs using a single 2D semiconductor. Logic inverters have been demonstrated using a variety of 2D materials including $MoS_2$, $MoTe_2$, $WSe_2$ and BP, but these prior demonstrations suffer from various shortcomings that limit their potential for use in high-performance complementary logic. For instance, unipolar inverters and ring oscillators consisting of depletion and enhancement mode $MoS_2$ n-MOSFETs[21,22] have been reported, but such circuits have high standby current and are not suitable for low-power applications. In addition, several demonstrations of complementary TMD circuits using a blanket substrate gate have been reported,[23-27] but such a device structure is not suitable for multi-stage circuits, which is essential for practical applications. In addition, in much of the prior work, the input and output voltage ranges of the inverters are not matched,[15,23-25,27,28] making them lack the capacity to drive subsequent logic inverter stages. In some cases, electrostatic doping has been utilized[25,26] to correct this problem, but once again, such individual device tuning is not

Symmetric Complementary Logic Inverter Using Integrated Black Phosphorus and MoS$_2$ Transistors

practical for large-scale implementation. Reports of WSe$_2$ logic inverters have been reported, but these circuits operated at low drive currents and the ability to simultaneously achieve high drive currents for both p- and n-MOSFETs in WSe$_2$ is unclear.[29-31] A promising alternative to using a single 2D material for CMOS circuits would be to combine MoS$_2$ n-MOSFETs with BP p-MOSFETs, and one such demonstration of a CMOS logic inverter has been reported in the literature.[15] However, the demonstration in reference 15 is not suitable for use in realistic logic circuits, once again due to poor input/output voltage window matching.

In this letter, we provide the first demonstration of a high-performance 2D logic inverter fabricated using an MoS$_2$ n-MOSFET and a BP p-MOSFET. The devices utilize local backgate electrodes with thin (20 nm) HfO$_2$ gate dielectrics. It is a true four-terminal device (IN, OUT, $V_{DD}$ and GND), in that it does not utilize any extrinsic biasing electrodes to shift the threshold voltages of the individual devices. Most importantly, the circuit provides voltage gain within a voltage window suitable for driving a subsequent logic stages and such gain is demonstrated for a supply voltage, $V_{DD}$, as low as 0.5 V. The circuit also provides stable performance at elevated temperatures typical of actual device operating conditions. Finally, the AC large and small signal operation is analyzed and operation at frequencies up to 100 kHz is demonstrated, which is the highest speed reported to date for a 2D CMOS logic circuit. These results establish a critical step towards creating high-performance logic circuits using 2D semiconducting materials.

## 2. Fabrication

The device fabrication started by using a bulk silicon wafer upon which a 110-nm-thick SiO$_2$ film was grown using thermal oxidation. After patterning alignment marks, a local gate electrode was patterned using electron-beam lithography (EBL). A 2-μm wide gate stripe was patterned in PMMA and then a combination of dry and wet etching was used to recess the SiO$_2$ before evaporating and lifting off Ti/Pd (10/40 nm) to from a quasi-planarized gate contact. Next, 20 nm of HfO$_2$ was deposited at 300 °C by atomic layer deposition (ALD), and this film served as the gate dielectric for the transistors. An MoS$_2$

# Symmetric Complementary Logic Inverter Using Integrated Black Phosphorus and MoS₂ Transistors

flake and a BP flake were then exfoliated and transferred onto the same gate finger using an optical aligning system. Afterward, EBL was again used to define the source and drain openings of both the MoS₂ n-MOSFET and the BP p-MOSFET. Finally, Ti/Au (10/80 nm) metallization was evaporated and lift-off, completing the fabrication process. The buried gate electrode served as the input terminal for the inverter, while the shared drain contact of the two devices served as the output electrode. The source contacts of the MoS₂ n-MOSFET and the BP p-MOSFET served as the ground (GND) and supply ($V_{DD}$) terminals of the inverter, respectively. Figures 1a-b, shows a schematic and circuit diagram of the inverter, while an optical micrograph of the completed circuit is shown in Figure 1c. The effective gate length, $L_{eff}$, (defined by the source-to-drain spacing) of both devices was 500 nm. The gate width, $W_g$, was defined by the flake dimensions and the BP p-MOSFET had $W_g = 16$ μm while the MoS₂ n-MOSFET had $W_g = 10$ μm.

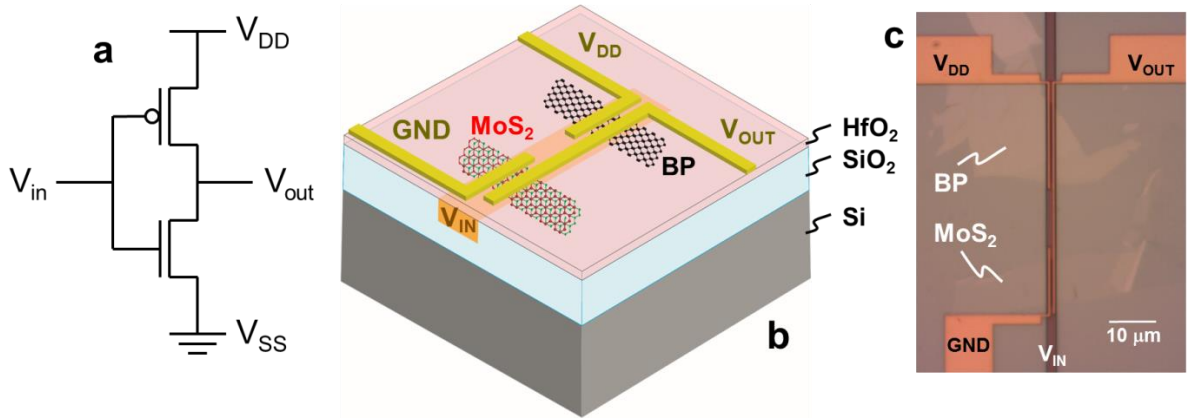

Figure 1. Illustration of the 2D material inverter. (a) Circuit schematic of CMOS logic inverter. (b) Schematic illustration of the integrated BP p-MOSFET and MoS₂ n-MOSFET. The devices are fabricated using a common buried gate electrode with thin HfO₂ gate dielectric. The source-to-drain spacings for both devices are 0.5 μm and the HfO₂ dielectric layer thickness is 20 nm. The widths of the MoS₂ and BP layers are 10 μm and 16 μm, respectively. (c) Optical micrograph of the inverter before deposition of the final contact layer.

## 3. Results and Discussion

The devices were measured with an Agilent B1500A semiconductor parameter analyzer in the dark using a cryogenic vacuum probe station at a pressure of $< 10^{-5}$ Torr. No surface passivation was utilized

Symmetric Complementary Logic Inverter Using Integrated Black Phosphorus and MoS$_2$ Transistors

and the initial characterization was performed at room temperature. Characterization was initially carried out separately for both the MoS$_2$ n-MOSFET and BP p-MOSFET and results are shown in Figure 2. In Figure 2a (Figure 2b), the drain current, $I_D$ of the MoS$_2$ n-MOSFET (BP p-MOSFET) is plotted vs. the drain-to-source voltage, $V_{DS}$, for terminal voltages between 0 and +2.5 V (0 and −2.5 V). The results show that the width-scaled drive current is well matched between the n- and p-MOSFETs, with on-current of ~ 50 µA/µm for both devices at $|V_{DS}| = |V_{GS}| = 2.5$ V. Transfer characteristics for both devices were also measured and the results are shown in Figs 2c and 2d. Here, $V_{GS}$ was swept between −1.5 and +1.5V and therefore, in order to limit the maximum terminal voltages applied to the device, a maximum value of $|V_{DS}| = 1.5$ V was used. While both devices are slightly depletion-mode, they display very symmetric linear threshold voltages, $V_T$, with the n-MOSFET (p-MOSFET) having $V_T = -0.8$ V (+0.8 V). The field-effect electron and hole mobilities were extracted from the linear $I_D$ vs. $V_{GS}$ characteristics and found to be 4.5 cm$^2$/Vs an 21 cm$^2$/Vs for the MoS$_2$ and BP, respectively, where we note that contact resistance was not subtracted from the measurements. In Figure 2c, for the MoS$_2$ n-MOSFET, an on-off current ratio > 10$^8$ was observed with a nearly-ideal subthreshold swing of 70 mV/decade (73 mV/decade) at $V_{DS} = +0.1$ V (+1.5 V). These results suggest that the buried gate electrode design provides a very high quality interface between MoS$_2$ and HfO$_2$. For the BP p-MOSFET, an on-off current ratio of ~ 10$^3$ was obtained and the devices show strong p-type behavior. The devices also showed poorer subthreshold slope compared to the MoS$_2$ devices and also slightly higher hysteresis, suggesting the possibility of some trapped moisture between the BP and HfO$_2$. Nevertheless, owing to its higher mobility, the BP p-MOSFET had a peak saturated transconductance, $g_m$, of 41 µS/µm at $V_{DS} = -1.5$ V while the MoS$_2$ n-MOSFET had a peak $g_m$ of 16 µS/µm at $V_{DS} = +1.5$ V.

Symmetric Complementary Logic Inverter Using Integrated Black Phosphorus and MoS$_2$ Transistors

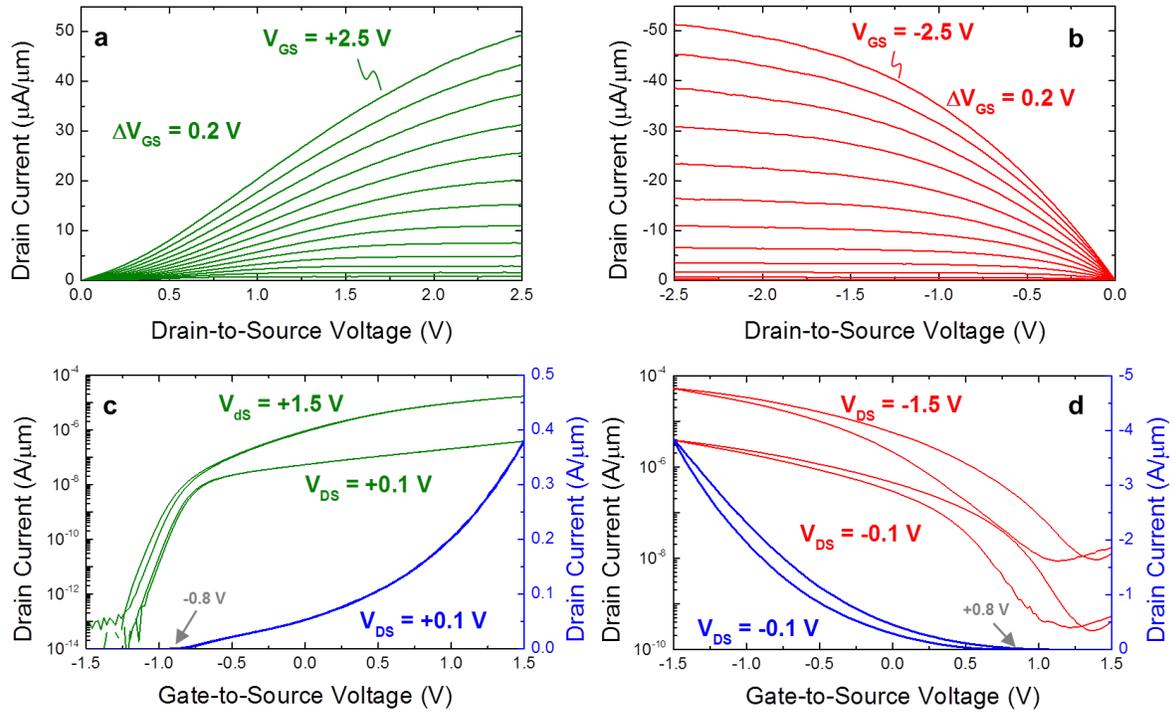

Figure 2. Characteristics of the individual devices of the logic inverter. (a) Drain current, $I_D$, vs. drain-to-source voltage, $V_{DS}$, characteristic for MoS$_2$ n-MOSFET. (b) $I_D$ vs. $V_{DS}$ characteristic for (BP) p-MOSFET. For both plots in (a) and (b), $|V_{DS}|$ was swept from 0 to 2.5 V and the maximum value of the absolute gate-to-source voltage, $|V_{GS}|$, was 2.5 V with a step of 0.2 V and $I_D$ was normalized by the widths of the individual devices. (c) $I_D$ vs. $V_{GS}$ characteristic of the MoS$_2$ n-MOSFET on both a semi-log (green) and linear (blue) scale. (d) $I_D$ vs. $V_{GS}$ characteristic of the BP p-MOSFET on both a semi-log (red) and linear (blue) scale. For the characteristics in (c) and (d), the values of $|V_{DS}|$ are 0.1 V and 1.5 V. Both directions of the gate voltage sweep are shown. The linear threshold voltages for n- and p-MOSFETs are approximately -0.8 V and +0.8 V, respectively.

In addition to testing the individual device elements, the integrated circuit was tested as an inverter and the results are shown in Figure 3. In all of these measurements, the common gate electrode was biased at an input voltage, $V_{IN}$, the shared drain contact was monitored as the output voltage, $V_{OUT}$, and the source contact of the BP p-MOSFET was biased at a fixed value of $V_{DD}$. Finally, all voltages were referred against the MoS$_2$ n-MOSFET source terminal voltage which was held at zero voltage source and is labeled GND in Figure 1. The inverter was tested at supply voltages ranging from $V_{DD}$ = 0.25 V to 2.5 V, in steps of 0.25 V. At each $V_{DD}$ value, $V_{OUT}$ and the inverter current were measured versus $V_{IN}$ between 0 V to $V_{DD}$. Figure 3a shows $V_{OUT}$ vs. $V_{IN}$, while the voltage gain is plotted vs. $V_{IN}$ in Figure 3b. Here, it can be observed that the device displays peak voltage gain, $G_{peak}$, great than 1 for supply $V_{DD} \geq$

Symmetric Complementary Logic Inverter Using Integrated Black Phosphorus and MoS$_2$ Transistors

0.5 V, while $G_{peak} > 2.5$ at $V_{DD} = 2.0$ V. A key feature of the gain characteristics is that the inversion and voltage gain are achieved in a symmetric input-output voltage window. Figure 3c shows the inverter current vs. input voltage for different supply voltages. The peak switching current is observed to have a peak for all voltage ranges, confirming the complementary nature of the circuit operation. Finally, in Figure 3d, the input-output and gain characteristics at $V_{DD} = 2.5$ V are plotted where the extracted noise margin has been extracted by creating a butterfly curve from the inverted input-output curve. This plot shows that the peak gain occurs at $V_{IN} = 1.2$ V, which is very close to the half the supply voltage of 2.5 V.

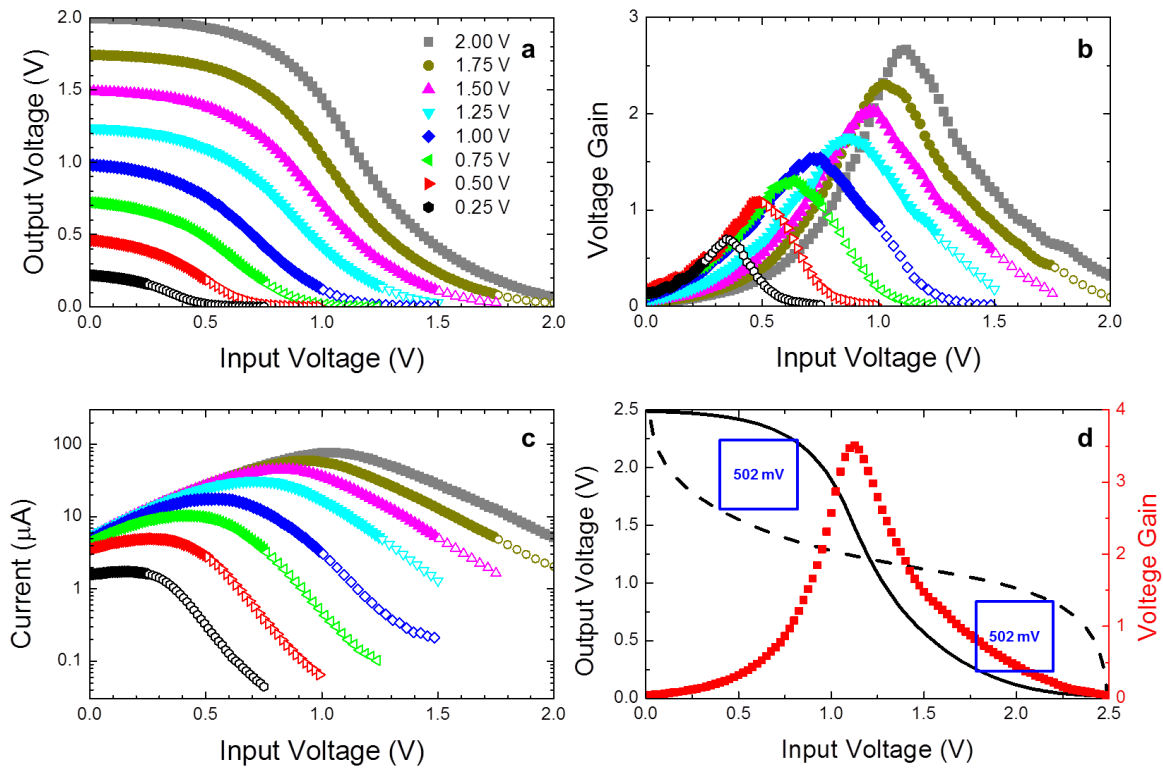

Figure 3. Results of measurements on integrated BP p-MOSFET / MoS$_2$ n-MOSFET logic inverter. (a) Output voltage, $V_{OUT}$, as a function of the input voltage, $V_{IN}$, at supply voltages, $V_{DD}$, ranging from 0.25 V to 2.0 V. The solid symbols indicate values of $V_{IN} \leq V_{DD}$, while the open symbols show points where $V_{IN} > V_{DD}$. (b) Voltage gain and (c) current vs. $V_{IN}$ for $V_{DD} = 0.25$ V to 2.0 V, where the symbol designations are the same as in (a). (d) Inverter $V_{OUT}$ (solid black line) and gain (red symbols) vs. $V_{IN}$ for inverter at $V_{DD} = 2.5$ V. The dashed black curve shows $V_{IN}$ vs. $V_{OUT}$, and the blue squares indicate the static noise margin of the inverter which is found to be > 500 mV.

Symmetric Complementary Logic Inverter Using Integrated Black Phosphorus and MoS$_2$ Transistors

The devices also have excellent noise margin, with a value of 502 mV extracted from the open area of the butterfly curve as shown in Figure 3d. These results indicate that our hybrid BP/MoS$_2$ inverter is capable of driving subsequent inverter stages. The $V_{DD}$-dependence of the inverter operation is summarized in Figure 4. As shown in Figure 4a, $G_{peak}$ increases from 1.1 to 3.5 going from $V_{DD}$ = 0.5 to 2.5 V, while the max-to-main current ratio (Figure 4b) also increases with increasing $V_{DD}$, where the delayed onset of the increase is due to the slightly non-linear turn-on of the MoS$_2$ n-MOSFET. Finally, the static noise margin vs. $V_{DD}$ is plotted in Figure 4c, where it can be seen that open butterfly characteristics are observed down to $V_{DD}$ = 1 V. Once again, reducing the on resistance of the n-MOSFET should allow inverter operation to even low supply voltages.

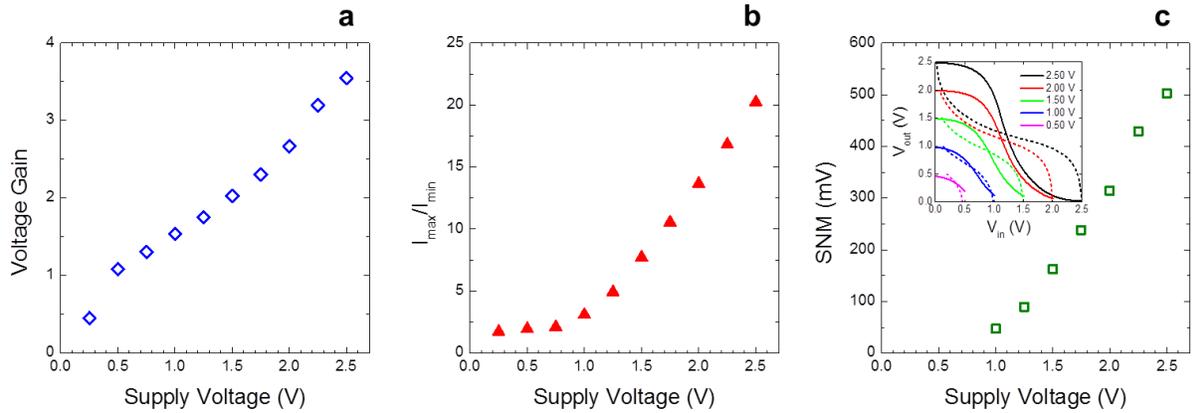

Figure 4. Dependence of room-temperature inverter characteristics on supply voltage, $V_{DD}$. (a) Maximum voltage gain of the inverter vs. $V_{DD}$ for conditions where $V_{IN} \leq V_{DD}$. (b) Ratio of maximum to minimum inverter current vs. $V_{DD}$, where the minimum current is determined from the average current between $V_{DD}$ = 0 and 2.5 V. (c) Static noise margin (SNM) vs. $V_{DD}$ where the SNM value is extracted using the butterfly curve method. Inset: butterfly curves for $V_{DD}$ = 0.50 to 2.5 V. Open butterfly characteristics are observed down to $V_{DD}$ = 1 V.

In order to evaluate the performance of the inverter in a realistic VLSI chip, the temperature dependence of the integrated inverter circuit was studied as a function of temperature which was varied between 270 K and 340 K. Figure 5a shows the in/out characteristics of the inverter, in which $V_{OUT}$ is plotted vs $V_{IN}$ with a supply voltage of 2.5 V, while the gain and drive current vs. $V_{IN}$ for the same temperatures are shown in Figures 5b and 5c. Several revealing trends are evident in the temperature-dependent data. First, the peak voltage gain is found to be virtually constant with temperature, and only a small decrease in the voltage of the peak gain characteristics is found with increasing temperature.

Symmetric Complementary Logic Inverter Using Integrated Black Phosphorus and MoS$_2$ Transistors

Secondly, the on-to-off current ratio decreases somewhat with increasing temperature and this is due to an increase in the gate-induced drain leakage (GIDL) of the BP p-MOSFET. Finally, it can be observed that when the input is high, the low output voltage tends to deviate from zero with higher temperature. This is also due to the off-state current flow in the p-MOSFET, which directly leads a non-zero voltage drop across the n-MOSFET, preventing the output low from reaching 0 V.

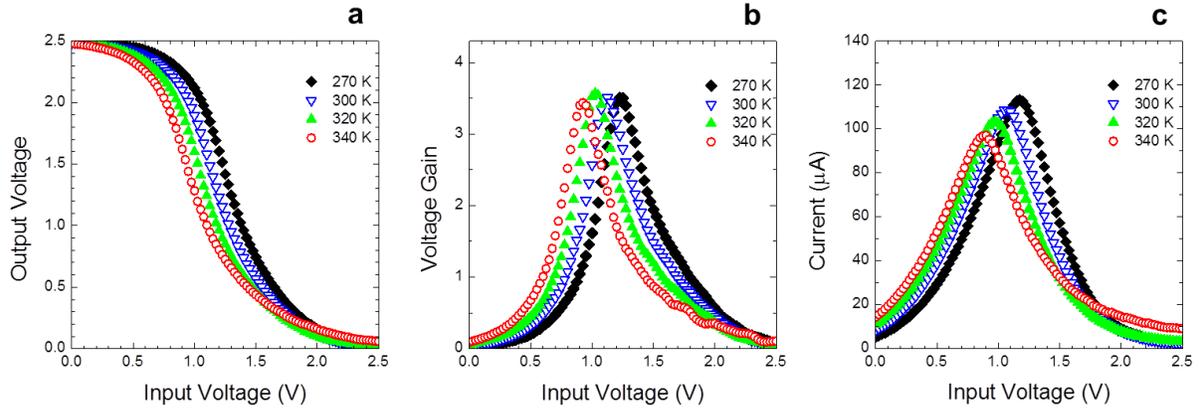

Figure 5. Temperature dependence of the inverter characteristics for temperatures ranging from 270 K to 340 K. (a) Temperature dependence of the $V_{OUT}$ vs. $V_{IN}$ characteristics. (b) Temperature dependence of voltage gain vs. $V_{IN}$ characteristics. (c) Temperature dependence of the inverter current vs. $V_{IN}$. For all measurements, the supply voltage, $V_{DD}$, is 2.5 V.

Finally, AC measurements were performed on the inverter and the results are shown in Figure 6. Here, a function generator was connected between the input gate electrode and ground and the resulting output signal between the shared drain terminal and ground was monitored using a digital oscilloscope. Figures 6a and 6b show $V_{IN}$ and $V_{OUT}$ vs. time at $V_{DD}$ = 2.5 V, were $f$ = 1 kHz and 100 kHz, for Figures 6a and 6b, respectively. The results show good logic operation up to 100 kHz, with the speed limited by parasitic capacitances and inductances associated with the vacuum probe station, with much higher speeds expected as will be described below. Small signal analysis was also performed on the devices and the results are shown in Figures 6c and 6d. Here, a sine-wave input was applied to the input terminals and the small-signal voltage gain, $G_{AC}$ was measured. In Figure 6c, $G_{AC}$ is plotted vs. the DC offset voltage, $V_{IN\text{-}DC}$, for different values of $V_{DD}$ ranging from 1 V to 2.5 V at a frequency of 1 kHz. Compared to the DC

Symmetric Complementary Logic Inverter Using Integrated Black Phosphorus and MoS$_2$ Transistors

characteristics performed at the same conditions, the peak AC gain was found to be larger, and this difference could be due to slow trapping effects in the gate dielectric, particularly in the BP p-MOSFET, as evidenced by the hysteretic behavior evident in Figure 2d. Figure 6d shows a plot of $G_{AC}$ vs. frequency at $V_{DD}$ = 2.5 V, there $V_{IN\text{-}DC}$ was adjusted to be at the peak gain condition. The gain roll-off occurs as expected with unity voltage gain reached at $f$ = 100 kHz.

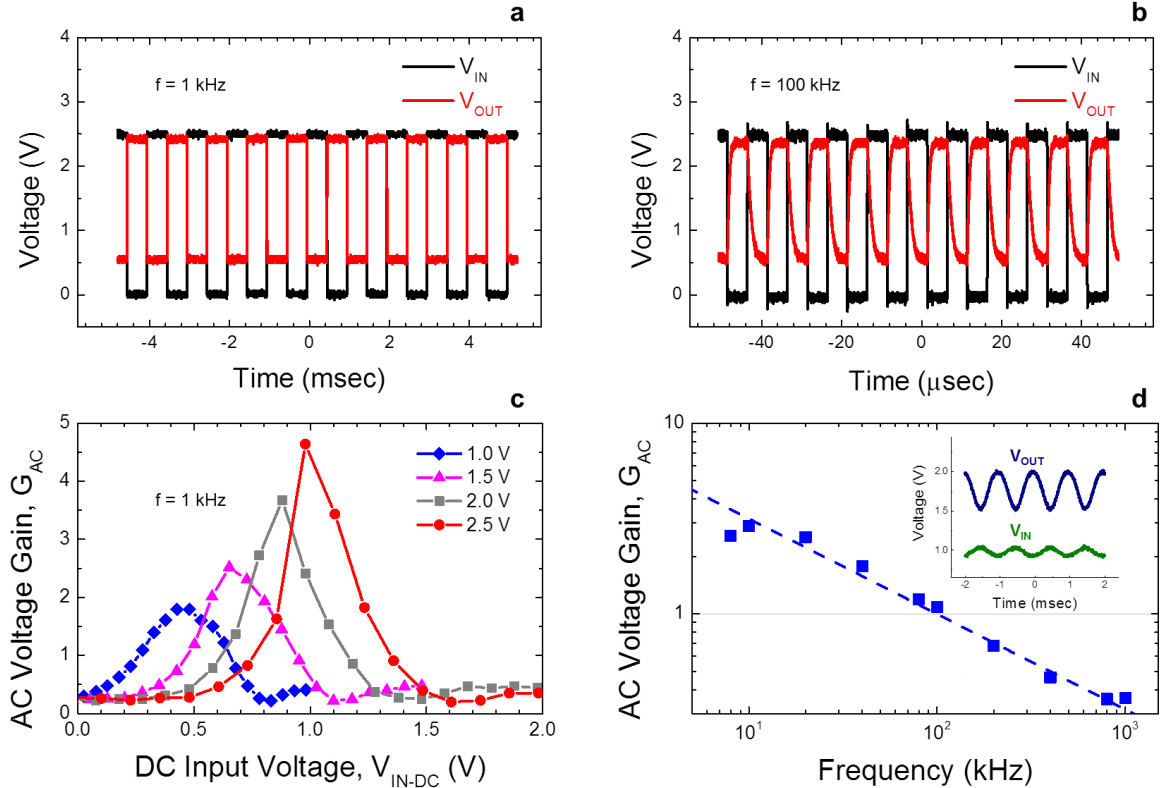

Figure 6. Digital and small-signal AC performance of inverter circuit. (a) Plot of input voltage, $V_{IN}$, and output voltage, $V_{OUT}$ vs. time at a frequency, $f$ = 1 kHz for inverter circuit. The supply voltage, $V_{DD}$ = 2.5 V and $V_{IN}$ was a square wave with minimum and maximum values of 0 and $V_{DD}$. (b) Same device as in (a) at $f$ = 100 kHz. (c) Plot of small signal AC voltage gain, $G_{AC}$, vs. DC input voltage, $V_{IN\text{-}DC}$ at $V_{DD}$ = 1.0 V (blue), 1.5 V (magenta), 2.0 V (grey) and 2.5 V (red). The input oscillator had a peak-to-peak voltage of 0.1 V and $f$ = 1 kHz. (d) Plot of $G_{AC}$ vs. $f$ at $V_{DD}$ = 2.5 V, where the device is biased near the peak gain point. The cutoff frequency is ~ 100 kHz, where the speed is dominated by parasitic capacitances associated with the test setup. Inset: Plot of $V_{IN}$ (green) and $V_{OUT}$ (blue) vs. time at $f$ = 1 kHz.

## 4. Discussion and Conclusions

The results reported in this paper provide important information on the suitability of integrating MoS$_2$ and BP to create CMOS logic circuits. First of all, the results show that these devices are capable of

Symmetric Complementary Logic Inverter Using Integrated Black Phosphorus and MoS$_2$ Transistors

reasonably symmetric performance, in that the current drive and threshold voltages are well matched, without the use of external control gates for threshold adjustment. In addition, the circuit performance is achieved using a common gate metal, dielectric and contact metallization, which could help to streamline future CMOS integration. In addition, while the exfoliation technique is not necessarily an extendable process, the results do show that the process of forming gate electrodes and subsequent transfer of the 2D material onto the pre-patterned substrate can result in high-performance devices and this work could help to spur development of techniques to transfer CVD material for more complex circuit operation.

It is clear that numerous improvements in the performance are possible with design optimization. In particular, while the threshold voltages are matched in our devices, they operate slightly in depletion mode, resulting in relatively high off-state leakage in our inverter. Thinning of the dielectric is likely to increase the threshold voltage in both devices. In addition, thinning the black phosphorus channel of the p-MOSFET should also help to improve the off-state leakage, since the increased band gap should have the effect of shifting the threshold voltage negative, as well as reducing the GIDL current. Improved matching could also be achieved by using a mesa etch to adjust the relative sizes of the MoS$_2$ and BP transistors.

Improved performance can also be achieved by eliminated parasitic elements. The intrinsic speed of the devices should be much faster than the current 100 kHz performance, which is attributed to both test setup capacitances and inductances as well as substrate coupling of the large probe pads. The intrinsic delay can be calculated as follows. The total capacitance, $C_{tot}$, of the combined circuit be calculated as $C_{tot} = LW \times \varepsilon_r \varepsilon_0 / t_{ox}$, where $L = 2$ μm is the total length of the gate electrode, $W = 26$ μm is the combined width of the p- and n-MOSFETs, $\varepsilon_r = 16.6$ is the dielectric constant of our ALD HfO$_2$ as determined from reference 20, $t_{ox} = 20$ nm and $\varepsilon_0$ is the permittivity of free space. These parameters produce a value of $C_{tot} = 0.38$ pF. Given the peak drive current, $I_{peak}$, of 108 μA at $V_{DD} = 2.5$ V, the intrinsic delay, $\tau$, can be calculated as $\tau = C_{tot} V_{DD} / I_{peak} = 8.8$ nsec, corresponding to a maximum frequency of 18 MHz.

Additional challenges that need to be addressed in order to improve the performance include improving the contact resistance, scaling the dielectric thickness and minimizing parasitic capacitance. From Figure 2, the contact resistance is very high, particularly in the MoS$_2$ device, and so reducing resistance arising from the Schottky contacts will be a key component in further enhancing the performance, particular at low supply voltages. Much more aggressive gate dielectric scaling should also be possible, particularly since 2D transistors with 7-nm HfO$_2$ have already been demonstrated.[20] In addition to allowing aggressive gate length scaling, dielectric scaling should also improve transmission coefficient of the contacts by increasing the electric field at the metal-semiconductor interface. Finally, utilization of a self-aligned geometry will be important in the future to minimize overlap capacitance between the gate and channel.

Finally, we note that some degradation in the device performance was observed over the course of the measurements reported in this paper. In particular, for the AC data in Figure 6, which was taken several weeks after the DC results, we found that the low input voltage was observed to be higher than in the DC shown in Figures 2-5. We believe this degradation is due to an increase in the GIDL current of the BP p-MOSFET, and is attributed to intermittent exposure to atmosphere between the DC and AC measurements. These results show that efficient passivation techniques will be needed for stable circuit operation, and the effect of passivation on the threshold voltages and current matching will be important aspects of future optimization.

In conclusion, we have demonstrated and characterized a complementary logic inverter fully based on integrated MoS$_2$ n-MOSFETs and BP p-MOSFETs. The devices utilize a backgate structure that allows high transconductance and excellent subthreshold slope to be obtained for both of the constituent transistors. The large switching current, symmetric input/output characteristics and high static margin of the inverter show the potential of these materials for use in large scale integrated circuits. The inverter also shows consistent operation over a wide temperature range and voltage gain at frequencies up to 100 kHz. Calculations suggest intrinsic speed > 10 MHz, and assuming CMOS scaling trends can be applied to 2D materials, then scaling the logic speed into the GHz regime should ultimately be possible.

Symmetric Complementary Logic Inverter Using Integrated Black Phosphorus and MoS$_2$ Transistors

These results are encouraging for creating high-performance logic circuits using hybrid integration of different 2D semiconducting materials.

**Acknowledgements**

This work was partially supported by the Defense Threat Reduction Agency Basic Research Award No. HDTRA1-14-1-0042, the National Science Foundation (NSF) under grant No. ECCS-1102278, and the NSF through the University of Minnesota MRSEC under Award No. DMR-1420013.

**Methods**

*Device and integrated circuit fabrication*. The fabrication of our inverter circuit started with dry thermal oxidation of a 110-nm-thick SiO$_2$ layer on a Si substrate. Alignment marks were first patterned on the substrate by spinning poly methyl methacrylate) (950 ka.u. PMMA) and then patterning with electron-beam lithography (EBL) using a Vistec EBPG 5000+ system. After development in 1:3 MIBK:IPA and rinsing in IPA; Ti / Au (10 nm / 100 nm) was deposited using electron-beam evaporation followed by a solvent liftoff in acetone followed by an IPA rinse. Next; the local back gate contacts were patterned. Once again; 950 ka.u. PMMA was spin-coated on the wafer and EBL was used to pattern 2-μm wide; 80-μm long stripes connected to enlarged pad regions for wafer probing. After development in 1:3 MIBK:IPA; the sample went through a 5sec oxygen plasma to remove PMMA residues and a reactive ion etching with CHF$_3$/CF$_4$/Ar to create a 40-nm deep recess in the SiO$_2$ layer. The sample was then etched in a 1:10 buffer oxide etch (BOE) for 12 seconds to create a roughly 50-nm recess in the SiO$_2$; and the recess depth was determined using a surface profilometer (KLA-Tencor P-7) before Ti / Pd (10 nm / 40 nm) was evaporated using electron-beam evaporation. After lift off in acetone / IPA; 20 nm of HfO$_2$ was deposited using atomic layer deposition (ALD) using Tetrakis(dimethylamido) hafnium(IV) and water vapor as the precursors. MoS$_2$ purchased from SPI and black phosphorus (BP) purchased from Smart Elements were then mechanically exfoliated onto Polydimethylsiloxane (PDMS) stamps activated on glass slides. With a specially designed optical alignment station; few-layer MoS$_2$ and BP flakes were aligned and transferred onto the same gate finger on the HfO$_2$-coated substrate. Atomic force microscopy

Symmetric Complementary Logic Inverter Using Integrated Black Phosphorus and MoS$_2$ Transistors

analysis after device fabrication revealed that both the MoS$_2$ and BP flakes had thicknesses of $8 \pm 1$ nm, where the thickness resolution was limited by roughness of the underlying gate metal. Due to degradation issues with the black phosphorus; the MoS$_2$ flake was exfoliated and aligned first; followed by the BP flake. A solvent clean was performed to remove PDMS residue and then PMMA was spin-coated right after transferring to prevent air degradation of the BP; and the sample was then stored in a black jar filled with desiccant. PMMA was spin-coated and EBL was then performed to open source and drain contact windows where the two devices shared the same drain contact as shown in Figure 1. Ti / Au (10 nm / 80 nm) metallization was again evaporated and lifted-off in acetone/IPA to complete the circuit fabrication. After completion of the lift off; the sample was loaded into the vacuum chamber of the Lakeshore cryogenic probe station for testing. No surface passivation was utilized.

*Device and circuit characterization.* All of the device and circuit characterization was performed under vacuum conditions ($\sim 10^{-5}$ Torr) using a Lakeshore CPX-VF cryogenic probe station with triaxial probe feedthrough connections. For the DC measurements, and Agilent B1500A semiconductor device parameter analyzer was utilized. The temperature-dependent measurements were performed with liquid nitrogen cooling and a heated stage feedback loop to stabilize the temperature. AC measurements were performed using a two-channel Keysight 33522B function generator, where one channel was used to supply the square wave or sinusoidal input waveform, while the other was used to provide the constant DC supply voltage. Both the input and output waveform data were monitored with a two-channel Keysight 3012C digital oscilloscope. For the input waveforms shown in Figures 6a and 6b, a correction was applied to the data to account for a zero offset in the oscilloscope calibration. All the testing channels and the circuit shared a same ground terminal.